\documentclass[a4paper,12pt]{article}
\usepackage{graphicx}
\usepackage{latexsym}
\usepackage{amsmath,amsthm}

\title{\bf
A transformed rational
function method and exact solutions to the $3+1$ dimensional Jimbo-Miwa equation
}
\date{}

\author{
Wen-Xiu Ma${}^{a}$\thanks{Email: {\tt mawx@cas.usf.edu} \ \  On
sabbatical leave of absence from University of South Florida, Tampa,
FL 33620, USA}{$\ $}  and Jyh-Hao Lee${}^{b}$\footnote{Email: {\tt
leejh@math.sinica.edu.tw}}{} \vspace{2mm}
\\
{\small {${}^{a}$}Department of Mathematics, Zhejiang Normal
University, Jinhua 321004, P.R. China}
\\
{ \small ${}^{b}$Institute of Mathematics, Academia Sinica, Taipei
11529, Taiwan} }

\begin{document}

\maketitle


\setlength{\baselineskip}{16.5pt}

\numberwithin{equation}{section}

\begin{abstract}

A direct approach to exact solutions of nonlinear partial differential equations
is proposed, by using rational function transformations.
 The new method
provides a more systematical and convenient handling of the solution process of nonlinear
 equations,
unifying
the tanh-function type methods,
the homogeneous balance method, the exp-function method, the mapping method, and the $F$-expansion type methods.
  Its key point is to search for rational solutions to variable-coefficient ordinary differential equations transformed from given partial differential equations.
 As an application,
the construction problem of exact solutions to
 the $3+1$ dimensional Jimbo-Miwa equation is treated, together with a B\"acklund transformation.


\end{abstract}


\vskip 0.5cm

\def \be {\begin{equation}}
\def \ee {\end{equation}}
\def \bea {\begin{eqnarray}}
\def \eea {\end{eqnarray}}
\def \ba {\begin{array}}
\def \ea {\end{array}}
\def \D {\displaystyle }
\newcommand{\R}{\mathbb{R}}

\newtheorem{theorem}{Theorem}[section]
\newtheorem{lemma}{Lemma}[section]

\def\cdot{{\scriptstyle\,\bullet\,}}

\section{\bf Introduction}

Partial differential equations (PDEs) describe various nonlinear
phenomena in natural and applied sciences such as fluid dynamics,
plasma physics, solid state physics, optical fibers, acoustics,
mechanics, biology and mathematical finance. Their solution spaces
are infinite-dimensional and contain diverse solution structures. It
is of significant importance to solve nonlinear PDEs from both
theoretical and practical points of view. Due to the nonlinearity of
differential equations and the high dimension of space variables, it
is a difficult job for us to determine whatever exact solutions to
nonlinear PDEs.

It has been a successful idea to generate exact solutions of
nonlinear wave equations by reducing PDEs into ordinary differential
equations (ODEs). Many approaches to exact solutions in the
literature follow such an idea, which contain the tanh-function
method \cite{LanW-JPA1990}-\cite{LiL-CPC2002}, the sech-function
method \cite{Ma-PLA1993}-\cite{ParkesZDH-PLA1998},
 the homogeneous balance method \cite{Wang-PLA1995,Wang-PLA1996}
the extended tanh-function
 method \cite{MaF-IJNM1996}-\cite{HanZ-ADE2000}, the sine-cosine method \cite{Yan-PLA1996}, the tanh-coth method \cite{Wazwaz-AMC2007},  the Jacobi elliptic function method \cite{LiuFLZ-PLA2001}, the exp-function method \cite{HeW-CSF2006}, the $F$-expansion method \cite{ZhouWW-PLA2003}, the mapping method \cite{Peng-CJP2003}, and the extended $F$-expansion method \cite{LiuY-CSF2004,ChenY-CSF2005}.
Given an ODE of differential polynomial type, either
constant-coefficient or variable-coefficient, one can always adopt
computer algebra systems to search for rational solutions pretty
systematically. This is one of the main reasons why those reduction
methods work well.

Based on this observation, we would, in this paper, like to propose
a direct and systematical approach to exact solutions of nonlinear
equations by using rational function transformations. The method is
very suitable for an easier and more effective handling of the
solution process of nonlinear equations, unifying the existing
solution methods mentioned above. Its key point is to find rational
solutions to variable-coefficient ODEs transformed from given
nonlinear PDEs. Together with an auto-B\"acklund transformation, an
application of our approach will generate new exact solutions to the
$3+1$ dimensional Jimbo-Miwa equation \cite{JimboM-PRIMS1983}
(called the Jimbo-Miwa equation in \cite{DorizziGRW-JMP1986}):
\begin{equation} P_{JM}(u):=u_{xxxy}+3u_yu_{xx}+3u_xu_{xy}+2u_{yt}-3u_{xz}=0
\label{eq:JM:pma-taipei-JM-2008}\end{equation}
where $u=u(x,y,z,t),\ u_x=\frac {\partial u}{\partial x},$ etc.

The $3+1$ dimensional Jimbo-Miwa equation
\eqref{eq:JM:pma-taipei-JM-2008} is
the second member in the entire Kadomtsev-Petviashvili hierarchy \cite{JimboM-PRIMS1983},
originally defined by a Hirota bilinear equation
\begin{equation}
[(D_x^3+2D_t)D_y-3D_xD_z)]\tau \cdot \tau =0,
\end{equation}
with the link being $u=2(\ln \tau )_x$.
The Jimbo-Miwa equation \eqref{eq:JM:pma-taipei-JM-2008}
passes the Painlev\'e test only for a subclass of solutions \cite{DorizziGRW-JMP1986} and
its symmetry algebra does not have a Kac-Moody-Virasoro structure \cite{RubinW-JMP1990}.
Nevertheless, different type solutions to
the Jimbo-Miwa equation \eqref{eq:JM:pma-taipei-JM-2008} are found (see, say,
\cite{TianG-CPC1996}-\cite{Wazwaz-AMC2008}).
The Hirota perturbation technique yields one- and two-soliton
solutions \cite{DorizziGRW-JMP1986} and dromion type solutions \cite{Wazwaz-AMC2008b}, and
the truncated Painlev\'e series leads to
a special B\"acklund transformation \cite{LiuJ-AMC2004}.
Obviously, the Jimbo-Miwa equation \eqref{eq:JM:pma-taipei-JM-2008}
has the following $x$- or $y$-independent solutions:
\begin{equation}u=f(x,t)+h(z,t),\ u=g(y,z)+h(z,t),\end{equation}
where $f$, $g$ and $h$ can be arbitrary functions in the indicated variables. These solutions contain more
special solutions: $u=f(x,t),\ u=g(y,z)$ and $ u=h(z,t)$, which are independent of
two variables of $x,y,z,t$.
 Starting with such solutions, various variable separated solutions
 have been presented by using the truncated Painlev\'e series \cite{LiuJ-AMC2004}
\cite{TangL-PLA2006} and abundant nonlinear coherent structures have
been exhibited \cite{TangL-PLA2006}. We will concentrate primarily
on constructing travelling wave solutions to the Jimbo-Miwa equation
\eqref{eq:JM:pma-taipei-JM-2008} by formulating a problem of finding
rational solutions to a transformed Jimbo-Miwa equation.

The paper is organized as follows. In Section
\ref{sec:Method:pma-taipei-JM-2008}, a unified formulation for
getting exact solutions to nonlinear equations is proposed, by using
rational function transformations. In Section
\ref{sec:Application:pma-taipei-JM-2008}, an application and an
auto-B\"acklund transformation are made to solve the $3+1$
dimensional Jimbo-Miwa equation \eqref{eq:JM:pma-taipei-JM-2008}. A
few of concluding remarks are given in the final section, along with
some polynomial solutions.

\section{A transformed rational function method}
\label{sec:Method:pma-taipei-JM-2008}

To describe our solution process, let us focus on a scalar $3+1$ dimensional partial differential equation
\begin{equation}
P(x,y,z,t,u_x,u_y,u_z,u_t,\cdots)=0,
\label{eq:PDE:pma-taipei-JM-2008}
\end{equation}
though the solution process also works for systems of nonlinear equations.
We assume that there are exact solutions to
the differential equation \eqref{eq:PDE:pma-taipei-JM-2008}:
\begin{equation}u(x,y,z,t)=u(\xi),\ \xi=\xi (x,y,z,t). \label{eq:TransformationtoXiEquation:pma-taipei-JM-2008}\end{equation}
Usually, we can have
 \begin{equation} \xi(x,y,z,t)=ax+by+cz-\omega t,\end{equation}
 where $a,b,c$ and $\omega$ are constants,
 in the constant-coefficient case, and
 \begin{equation} \xi(x,y,z,t)=a(t)x+b(t)y+c(t)z-\omega (t),\end{equation}
 where $a(t),b(t),c(t)$ and $\omega(t)$ are functions of $t$,
 in the $t$-dependent-coefficient case.
 Under the transformation \eqref{eq:TransformationtoXiEquation:pma-taipei-JM-2008},
 the partial differential equation \eqref{eq:PDE:pma-taipei-JM-2008} is put into an ordinary differential equation:
\begin{equation}
Q(x,y,z,t,u^{(r)},u^{(r+1)},\cdots)=0,
\label{eq:transformedPDE:pma-taipei-JM-2008}
\end{equation}
where $u^{(i)}=\frac {d^i u}{d\xi ^i},\ i\ge 1$, and $r$ is the least order of derivatives in the equation.
To keep the solution process as simple as possible, the function $Q$ should not be a total $\xi$-derivative of another function. Otherwise, taking integration with respect to $\xi$ further reduces the transformed equation.

An important step in the solution process is to introduce a new variable $\eta =\eta (\xi) $ by a solvable ordinary differential equation, for example, a first-order differential equation:
\begin{equation}
\eta ' =T=T(\xi,\eta ),
\label{eq:defofEta:pma-taipei-JM-2008}
\end{equation}
where  $T$ is a function of $\xi $ and $\eta $, and the prime denotes the derivative with respect to $\xi$.
 In case that we have a general second-order differential equation to begin with, we should first obtain its first integrals \cite{Feng-PLA2002} and then use the method of planar dynamical systems to solve \cite{LiWZ-JBCASE2006}.
Two simple solvable cases of the above function $T$ are as follows:
\begin{equation}
T=T(\eta )= \eta,\  T=T(\eta )=\alpha +\eta ^2 ,\ \alpha =\textrm{const.}
\label{eq:formsofEta:pma-taipei-JM-2008}
\end{equation}
The corresponding first-order equations have the particular solutions
$
\eta = e^\xi
$
and
\begin{equation}
\eta = \left\{\ba{l}
-\frac 1 \xi , \ \textrm{when}\ \alpha =0,\vspace{2mm}\\
-\sqrt{-\alpha }\,\tanh \sqrt {-\alpha }\, \xi\ \,\textrm{or}\ \,-\sqrt{-\alpha }\,\coth \sqrt {-\alpha }\, \xi,\ \textrm{when}\ \alpha <0,\vspace{2mm}\\
\sqrt{\alpha }\,\tan \sqrt {\alpha }\, \xi\ \,\textrm{or}\ \,-\sqrt{\alpha }\,\cot \sqrt {\alpha }\, \xi,\ \textrm{when}\ \alpha >0,
\ea \right.
\end{equation}
respectively \cite{MaF-IJNM1996}.
Those two cases correspond to the exp-function method and the extended tanh-function method, respectively.

More general assumptions than \eqref{eq:formsofEta:pma-taipei-JM-2008}
 can engender special function solutions to nonlinear wave equations. For instance,
taking $(\eta ' )^2=S(\eta )$ with some fourth-order polynomials $S(\eta )$ in $\eta $ (or equivalently,
   $\eta ''=R(\eta )$ with some third-order polynomials $R(\eta )$ in $\eta $)
 can yield Jacobi elliptic function solutions; and such assumptions are the bases for the extended tanh-function method,
 the $F$-expansion method and the extended $F$-expansion method, and work for many particular nonlinear wave equations.

The basic idea of using solvable ordinary differential equations was successfully used to solve the $2+1$ dimensional Korteweg-de Vries-Burgers equation in
\cite{Ma-JPA1993}, based on $a\eta ''+b\eta ' +c \eta ^2+d\eta =0$ ($a,b,c,d=\textrm{const.}$), and
the Kolmogorov-Petrovskii-Piskunov equation
in \cite{MaF-IJNM1996}, based on $\eta ' = 1\pm \eta ^2$. Later it was broadly adopted in the extended tanh-function method \cite{Fan-PLA2000,HanZ-ADE2000}, the tanh-coth method \cite{Wazwaz-AMC2007}, the $F$-expansion method \cite{ZhouWW-PLA2003}, the mapping method \cite{Peng-CJP2003,Yomba-CSF2004}, and the extended $F$-expansion method \cite{LiuY-CSF2004,ChenY-CSF2005}.

Let us proceed to consider rational functions
\begin{equation}
v(\eta )=\frac {p(\eta) }{q(\eta)}=\frac {p_m\eta ^m+p_{m-1}\eta ^{m-1}+\cdots  +p_0}{q_n\eta ^n+q_{n-1}\eta ^{n-1}+\cdots +q_0},
\label{eq:defofv:pma-taipei-JM-2008}
\end{equation}
where $m$ and $n$ are two natural numbers, and $p_i$, $0\le i\le m$ and $q_i,\ 0\le i\le n$ are normally constants but could be functions of the independent variables as in the $F$-expansion type method.
All Laurent polynomial and polynomial functions are only special examples of rational functions.
We search for travelling wave solutions determined by
\begin{equation}
u^{(r)}(\xi)=v(\eta )=\frac {p(\eta) }{q(\eta)},
\label{eq:formofu:pma-taipei-JM-2008}
\end{equation}
where $p(\eta)$ and $q(\eta )$ are polynomials as indicated above.
It is direct to compute that
\begin{equation}
u^{(r+1)} = T v' ,\ u^{(r+2)}= T\partial _\eta u^{(r)} =Tv''+T' v' ,\ \cdots,
\label{eq:derivativesofuwitheta:pma-taipei-JM-2008}
\end{equation}
which is based on $\partial _\xi =T\partial _\eta .$
Note that by the prime, we denote the derivatives with respect to the involved variable, for instance,  $u'=\frac {d u}{d\xi}$, $v'=\frac {dv}{d\eta}, $ and $v''=\frac {d^2v}{d\eta ^2}$.

Now we assume that the transformed equation
\eqref{eq:transformedPDE:pma-taipei-JM-2008} is a rational function equation of $\eta $
 with a given pair of $m$ and $n$. This can be achieved for all nonlinear equations of differential polynomial type,
when $T$ is, for example, a rational function in $\eta $.
Then we just need to force the numerator of the resulting rational function in the transformed equation
to be zero. This yields a system of algebraic equations on all variables $a,b,c,\omega$, $p_i,\ 0\le i\le m$ and $q_i,\ 0\le i\le n$,
and solve this system (which could be a differential system as in the $F$-expansion type method) to determine $p(\eta )$, $q(\eta)$ and $\xi$.
Finally, integrating $v(\eta)$ with respect to $\xi$, $r$ times,
we obtain a class of travelling wave solutions
\begin{eqnarray}
&& u(x,y,z,t)=u(\xi)= \underbrace{\int \cdots\int}_{r}\, \D \,\frac
{p(\eta(\xi))}{q(\eta (\xi ))}\,\, {d\xi\cdots d\xi }
\nonumber \\
&& := \int_{0}^\xi \int_0^{\xi_r} \cdots\int_{0}^{\xi_2}\, \D
\,\frac {p(\eta(\xi_1))}{q(\eta (\xi_1 ))}\,\, {d\xi_1\cdots
d\xi_{r-1}d\xi_r }
 +
 \sum_{i=1}^r d_i\xi ^{r-i}
 , \label{eq:expofmultipleindefiniteintegral:pma-taipei-JM-2008} \end{eqnarray}
where $d_i$, $1\le i\le r$, are arbitrary constants. If $r=1$, there
is only the last definite integral over $[0,\xi]$ in
\eqref{eq:expofmultipleindefiniteintegral:pma-taipei-JM-2008}.
  The resulting solutions will definitely contain a
polynomial part in $\xi$, when $r> 1$.

{\bf The case of the exp-function method:}

If we take $\eta =e ^\xi$, then the solution function is determined
by
\[u(\xi )=\underbrace{\int \cdots\int}_{r}\, \, \frac {p_m e^{m\xi} +p_{m-1} e ^{(m-1)\xi} +\cdots +p_0}{q_n e^{n\xi} +q_{n-1} e ^{(n-1)\xi}+ \cdots +q_0}\,\, {d\xi\cdots d\xi }.\]
This can yield solutions generated by the exp-function method \cite{HeW-CSF2006}.

{\bf The case of the extended tanh-function method:}

If we take $\eta = -\tanh (\xi) $, then the solution function is determined
by
\[u(\xi )=\underbrace{\int \cdots\int}_{r}\, \,\frac {(-1)^mp_m \tanh ^{m}(\xi) +(-1)^{m-1}p_{m-1} \tanh ^{m-1}(\xi)+ \cdots +p_0}{
(-1)^nq_n\tanh ^{n}(\xi) +(-1)^{n-1}q_{n-1} \tanh ^{n-1}(\xi)+ \cdots +q_0}\,\, {d\xi\cdots d\xi }.\]
This can yield solitary wave solutions.
If we take $\eta = \tan (\xi) $, then the solution function is determined
by
\[u(\xi )=\underbrace{\int \cdots\int}_{r}\, \,\frac {p_m \tan ^{m}(\xi) +p_{m-1} \tan ^{m-1}(\xi)+ \cdots +p_0}{q_n \tan ^{n}(\xi) +q_{n-1} \tan ^{n-1}(\xi)+ \cdots +q_0}\,\, {d\xi\cdots d\xi }.\]
This can yield periodic wave solutions.
The other choices of $\eta = -\coth (\xi) $ and $\eta =- \cot (\xi) $ generate similar type exact solutions. Note that we only use the selection of $\alpha =1$ and a general nonzero value of $\alpha $ may lead to more general solutions. The selection of $\alpha =0$ presents rational solutions. If $v$ is a polynomial and $r=0$, then $u$ gives an exact solution in the form of a finite series in $\tanh \xi$ or $\tan \xi$, which is obtainable by the tanh-function type method \cite{ParkesZDH-PLA1998,MaF-IJNM1996,Fan-PLA2000}.

The solution process described above unifies the existing methods using tanh-function type functions, tan-function type functions, the exponential functions and the Jacobi elliptic functions, and allows us to carry out the involved computation more systematically and conveniently by powerful computer algebra systems such as Maple, Mathematica, MuPAD and Matlab. While
applying to construction of special function solutions to nonlinear equations, we have to pay special attention to the particular forms of given nonlinear equations to get workable transformed equations.

\section{\bf Solving the $3+1$ dimensional Jimbo-Miwa equation}
\label{sec:Application:pma-taipei-JM-2008}

To generate travelling wave solutions \begin{equation}
u(x,y,z,t)=u(\xi),\ \xi=ax+by+cz-\omega t,\end{equation}
where $a,b,c,$ are the angular wave numbers and $\omega$ are the wave frequency,
 we only need to solve the reduced
 $3+1$ dimensional Jimbo-Miwa equation \eqref{eq:JM:pma-taipei-JM-2008}:
\begin{equation} a^3bu^{(4)} +6a^2bu'u'' -(2b\omega +3ac)u''=0,\end{equation}
where the prime denotes the derivatives with respect to $\xi $.
Integrate it once with respect to $\xi $ to obtain
\[ a^3bu''' +3a^2b(u') ^2 -(2b\omega +3ac)u'=0.\]
We set $r=1$ and $u'=v$, and then,
we have the transformed Jimbo-Miwa equation
\begin{equation}
{a}^{3}b T ^{2}v''  +{a}^{3}b TT'
  v' +3{a}^{2}b  v  ^{2}-(2b \omega  +3ac) v
=0,
\label{eq:transformedJM:pma-taipei-JM-2008}
\end{equation}
where the prime denotes the derivatives with respect to $\eta $.

\subsection{The case of $\eta '=\eta $}
\label{sec:subsectionofcase1:pma-taipei-JM-2008}

In this case, the transformed Jimbo-Miwa equation \eqref{eq:transformedJM:pma-taipei-JM-2008} becomes
\begin{equation}
{a}^{3}b \eta
^2  v'' +{a}^{3}b \eta  v '  + 3{a}^{2}b v^2-(2b
\omega  +3ac) v
=0.
\label{eq:case1oftransformedJM:pma-taipei-JM-2008}
\end{equation}

We try to search for a rational solution $v$ with $m=n=3$. A direct computation with Maple tells
that there are only two choices with a non-constant $v$ and $abc\ne 0$:
\begin{equation}
v(\eta )=\frac {4aq_1q_2\eta }{4q_2^2\eta ^2+4q_1q_2\eta +q_1^2},\ \omega =\frac{a(a^2b-3c)}{2b}
,
\label{eq:case1ofm=n=3:pma-taipei-JM-2008}
\end{equation}
and
\begin{equation}
v(\eta)=\frac {p_0(p_1^2\eta ^2+16 p_0p_1\eta +16p_0^2)}{q_0(p_1^2\eta ^2-8p_0p_1\eta +16 p_0^2)}
,\ a =-\frac {3p_0}{q_0},\ \omega =\frac {9p_0(3bp_0^2+cq_0^2)}{2bq_0^3}
.
\label{eq:case2ofm=n=3:pma-taipei-JM-2008}
\end{equation}
Accordingly, we have the following
 travelling wave solutions to the $3+1$ dimensional Jimbo-Miwa equation \eqref{eq:transformedJM:pma-taipei-JM-2008}:
\begin{equation}
u(x,y,z,t)=
-\frac {2aq_1}{q_1+2q_2 e^\xi }+d,\ \xi=ax+by+cz-\frac{a(a^2b-3c)}{2b} t,
\label{eq:solsofcase1ofm=n=3:pma-taipei-JM-2008}
\end{equation}
and
\begin{equation}
u(x,y,z,t)=\frac {24p_0^2}{q_0(4p_0-p_1e ^\xi )}+\frac {p_0}{q_0}\xi +d
,\ \xi =-\frac {3p_0}{q_0}x+by+cz-\frac {9p_0(3bp_0^2+cq_0^2)}{2bq_0^3}t
,
\label{eq:solsofcase2ofm=n=3:pma-taipei-JM-2008}
\end{equation}
where the involved constants are all arbitrary. Note that the second class of travelling wave solutions, defined by \eqref{eq:solsofcase2ofm=n=3:pma-taipei-JM-2008}, contain a linear function in $\xi$.

We point out that all exact solutions generated in \cite{OzisA-PLA2008} belong to the above first class of travelling wave solutions, defined by \eqref{eq:solsofcase1ofm=n=3:pma-taipei-JM-2008}, which can be observed by multiplying common factors, re-scaling
the involved constants and checking the rational form of the $v$-function. Only for the solution in the formula (30) of \cite{OzisA-PLA2008}, one first needs to cancel one common factor $e^\xi +b_1$.

\subsection{The case of $\eta '=\alpha +\eta ^2 $}

In this case, the transformed Jimbo-Miwa equation \eqref{eq:transformedJM:pma-taipei-JM-2008} becomes
\begin{equation}
a^3b\eta ^4 v''+ 2a^3b \eta ^3 v'+ 2{a}^{3}b \alpha \eta ^2  v ''  +{a}^{3}b \alpha ^2  v'' + 2{a}^{3}b\alpha \eta v' +3a^2bv^2 - (2b
\omega  +3ac) v
=0.
\label{eq:case2oftransformedJM:pma-taipei-JM-2008}
\end{equation}

We try to search for a rational solution $v$ with $m=3$ and $n=1$. Similarly, a direct computation with Maple tells
that there are only two choices with a non-constant $v$ and $abc\ne 0$:
\begin{equation}
v(\eta )=-2a\alpha -2a\eta ^2,\ \omega=-\frac {a(4ba^2\alpha +3c)}{2b}
,
\label{eq:case1ofm=3andn=1:pma-taipei-JM-2008}
\end{equation}
and
\begin{equation}
v(\eta )=
-\frac 2 3 a\alpha  -2a \eta ^2,\ \omega =\frac {a(4a^2b\alpha -3c)}{2b}
.\label{eq:case2ofm=3andn=1:pma-taipei-JM-2008}
\end{equation}

Through re-scaling the involved constants, we can know that the general nonzero
value of $\alpha $ does not engender more general solutions in the above two cases.
Therefore, we select $\alpha =1$ and take
$\eta =\tan \xi$ and $\eta =-\cot \xi$, we obtain
the corresponding travelling wave solutions to the $3+1$ dimensional Jimbo-Miwa equation \eqref{eq:transformedJM:pma-taipei-JM-2008}:
\begin{equation}
\left\{ \ba{l}
\D u(x,y,z,t)=-2a \tan \xi +d,\
\xi=ax+by+cz+\frac {a(4ba^2 +3c)}{2b}t
,
\vspace{2mm}\\
\D u(x,y,z,t)=-2a \tan \xi + \frac {4a \xi } 3 +d, \ \xi =ax+by+cz-\frac {a(4a^2b -3c)}{2b}t
,
\ea \right.
\label{eq:sol1ofm=3andn=1:pma-taipei-JM-2008}
\end{equation}
and
\begin{equation}
\left\{\ba{l}
\D u(x,y,z,t)=2a \cot \xi +d,\
\xi=ax+by+cz+\frac {a(4ba^2 +3c)}{2b}t
,
\vspace{2mm}\\
\D u(x,y,z,t)=2a \cot \xi + \frac {4a \xi } 3 +d, \ \xi =ax+by+cz-\frac {a(4a^2b-3c)}{2b}t
,
\ea \right.
\label{eq:sol2ofm=3andn=1:pma-taipei-JM-2008}
\end{equation}
where the involved constants are all arbitrary. Note that each second class of travelling wave solutions contain a linear function in $\xi$.
The selection of $\alpha =0$ leads to a class of rational solutions
\begin{equation}
u(x,y,z,t)=\frac {2a}\xi +d=\frac{4ab}{2abx+2b^2y+2bcz+3act}+d,
\end{equation}
where the involved constants are all arbitrary.

It is also direct to see that if we take
$\eta =- \tanh \xi$ and $\eta =-\coth  \xi$, we will
obtain the same travelling wave solutions as those presented in the sub-section \ref{sec:subsectionofcase1:pma-taipei-JM-2008}.
Our solutions generated from \eqref{eq:case1ofm=n=3:pma-taipei-JM-2008} and \eqref{eq:case1ofm=3andn=1:pma-taipei-JM-2008} contain the
solutions given in \cite{Senthilvelan-AMC2001}, and reduce to the solutions in the $2+1 $ dimensional case of $u=u(x,y,t)$ in \cite{ChenD-NA2005}.

\subsection{B\"acklund transformation}

Let $u=u(x,y,z,t)$ be a solution to the
 $3+1$ dimensional Jimbo-Miwa equation \eqref{eq:JM:pma-taipei-JM-2008}. Evidently, if a function $v=v(x,y,z,t)$ satisfies
\begin{equation}
 3u_yv_{xx}+3u_{xx}v_y+3u_{xy}v_x+3u_xv_{xy}+P_{JM}(v)=0,
 \label{eq:condofBT:pma-taipei-JM-2008}
\end{equation}
where $P_{JM}$ is the Jimbo-Miwa operator defined in \eqref{eq:JM:pma-taipei-JM-2008},
 then the sum of the two functions, $w=u+v$, gives another solution to the Jimbo-Miwa equation \eqref{eq:JM:pma-taipei-JM-2008}.
Therefore, once we find a function $v$ satisfying \eqref{eq:condofBT:pma-taipei-JM-2008}, we get a new solution $w=u+v$ from a known one $u$. This forms a general auto-B\"acklund transformation for the Jimbo-Miwa equation \eqref{eq:JM:pma-taipei-JM-2008}.

{\bf Sub-BT1:} First, it follows directly from the above B\"acklund transformation that
  if two solutions $u$ and $v$ of the Jimbo-Miwa equation \eqref{eq:JM:pma-taipei-JM-2008} satisfy
 \begin{equation}
 u_yv_{xx}+u_{xx}v_y+u_{xy}v_x+u_xv_{xy}=0,
 \label{eq:specialcondofBT:pma-taipei-JM-2008}
\end{equation}
then $w=u+v$ is a third solution to the Jimbo-Miwa equation \eqref{eq:JM:pma-taipei-JM-2008}.
Further, we easily see that any solution $u=u(x,y,z,t)$ plus an arbitrary function $h(z,t)$ gives a new solution for the Jimbo-Miwa equation \eqref{eq:JM:pma-taipei-JM-2008}.

{\bf Sub-BT2:} Second, if we take \begin{equation}v=2(\ln \phi)_x=2\frac {\phi_x}{\phi}, \end{equation}
then $w=u+2(\ln \phi)_x$ solves the Jimbo-Miwa equation \eqref{eq:JM:pma-taipei-JM-2008}, when
$\phi=\phi(x,y,z,t)$ satisfies
  \begin{equation}
\left\{ \ba {l}
\phi_{xxxy}+3u_x\phi_{xy}+3u_y\phi_{xx}+2\phi_{yt}-3\phi_{xz}=0,
\vspace{2mm}\\
\phi_{xxx}\phi_y+ 3\phi_{x}\phi_{xxy}-3\phi_{xx}\phi_{xy}+3u_x\phi_{x}\phi_{y}+3u_y\phi_{x}^2
+2\phi_{y}
\phi_{t}-3\phi_{x}\phi_{z}=0,
\ea \right.
\label{eq:condofBTofLiu:pma-taipei-JM-2008}
\end{equation}
where $u$ is a solution to the Jimbo-Miwa equation \eqref{eq:JM:pma-taipei-JM-2008}.
 This special B\"acklund transformation was introduced through the truncated Painlev\'e series
  for the Jimbo-Miwa equation \eqref{eq:JM:pma-taipei-JM-2008} in \cite{LiuJ-AMC2004} and for the derivative Jimbo-Miwa equation $(P_{JM}(u))_x=0$ in \cite{TianGH-CMA2002}. It was successfully used to present diverse variable separated solutions involving arbitrary functions of two variables for the Jimbo-Miwa equation in \cite{LiuJ-AMC2004} \cite{TangL-PLA2006}.

One one hand, if we take a solution $u=f(x,t)+h(z,t)$ and
\begin{equation}
\phi=a_0+a_1p(\frac 3 {c_0}y+z)+a_2q(x,t)+a_3p(\frac 3 {c_0}y+z)q(x,t),
\end{equation}
where $ a_i,\ 0\le i\le 3$, and $c_0$ are constants,
then the two conditions in \eqref{eq:condofBTofLiu:pma-taipei-JM-2008} will be satisfied when
 \begin{equation}
2q_t+q_{xxx}-q_x(c_0-3f_x)=0,
\end{equation}
which determines the function $f$. All the involved constants and the functions $h,p,q$ are arbitrary.
This generates the class of exact solutions to the Jimbo-Miwa equation \eqref{eq:JM:pma-taipei-JM-2008}, presented in \cite{TangL-PLA2006}.
More specially, when $a_1=a_2=0$, we obtain a sub-class of solutions given in \cite{LiuJ-AMC2004}.

On the other hand, if we take a solution $u=g(y,z)+h(z,t)$ and
\begin{equation}
\phi=a_1+a_2p(y,z)q(x,t),
\end{equation}
where $ a_1$ and $a_2$ are constants,
then the two conditions in \eqref{eq:condofBTofLiu:pma-taipei-JM-2008} will be satisfied when \cite{LiuJ-AMC2004}
 \begin{equation}
(q_x)^2-qq_xx=0,\ p_y(q_{xxx}+2q_t) +3 g_y pq_{xx}  -3 p_zq_x=0.
\end{equation}
A particular case is given by
\begin{equation}
q(x,t)=e^{c_1x+c_2t},\ g(y,z)=-\frac 1{3c_1^2}\int \frac {(c_1^3+2c_2)p_y-3c_1p_z}{p}\,dy,
\end{equation}
where $ c_1$ and $c_2$ are constants,
and thus the resulting new solution reads
\begin{equation}
w= \frac {2 a_2 c_1p(y,z) e^{c_1x+c_2t} }{a_1+a_2 p(y,z) e^{c_1x+c_2t} }
-\frac 1{3c_1^2}\int \frac {(c_1^3+2c_2)p_y-3c_1p_z}{p}\,dy +h(z,t),
\end{equation}
where all the involved constants and functions are all arbitrary.

{\bf Sub-BT3:} Third, if we take a travelling wave solution $u=u(x,y,z,t)=
u(ax+by+cz-\omega t)$ to the Jimbo-Miwa equation \eqref{eq:JM:pma-taipei-JM-2008}, then the function
\begin{equation}
w(x,y,z,t)=u(ax+by+cz-\omega t)+ a'x+b'y+c'z-\omega ' t +d',
\label{eq:sBT:pma-taipei-JM-2008}
\end{equation}
where $a',b',c',d',\omega'$ are all constants satisfying \begin{equation}ab'+a'b=0,\end{equation}
presents a new solution to the Jimbo-Miwa equation \eqref{eq:JM:pma-taipei-JM-2008}.

Now taking advantage of this feature in \eqref{eq:sBT:pma-taipei-JM-2008},
it is direct to check by multiplying common factors and re-scaling
the involved constants, particularly the angular wave numbers and the wave frequency, that all exact solutions presented in \cite{Wazwaz-AMC2008} are special examples of combination solutions of the travelling wave solutions, defined in \eqref{eq:solsofcase1ofm=n=3:pma-taipei-JM-2008}, \eqref{eq:solsofcase2ofm=n=3:pma-taipei-JM-2008},
 \eqref{eq:sol1ofm=3andn=1:pma-taipei-JM-2008} and \eqref{eq:sol2ofm=3andn=1:pma-taipei-JM-2008}, and some linear function solutions.

\section{Concluding remarks}
\label{sec:Concludingremarks:pma-taipei-JM-2008}

A new systematical solution procedure to constructing exact solutions to nonlinear partial differential equations, both constant-coefficient  and variable-coefficient, is proposed, based upon rational function transformations.
 Its key point is to search for rational solutions to variable-coefficient ordinary differential equations transformed from given partial differential equations.
Together with an auto-B\"acklund transformation, an application of our method
leads to various new travelling wave solutions to
 the $3+1$ dimensional Jimbo-Miwa equation \eqref{eq:JM:pma-taipei-JM-2008}.

 Our method
 allows us carry out the solution process of nonlinear wave
 equations more systematically and conveniently by computer algebra systems such as Maple, Mathematica, MuPAD and Matlab,
unifying
the tanh-function method, the sech-function method, the homogeneous balance method, the extended tanh-function
 method, the sine-cosine method, the tanh-coth method, the Jacobi elliptic function method, the exp-function method, the $F$-expansion method, the mapping method and the extended $F$-expansion method.
The presented method
also works for systems of nonlinear wave equations, and
it can be applied to construction of special function solutions, like
the extended $F$-expansion method \cite{LiuY-CSF2004,ChenY-CSF2005}, the extended mapping method \cite{WakilA-PLA2006} and the improved extended $F$-expansion method \cite{WangZ-CSF2005} \cite{ZhangX-AMC2006} \cite{Abdou-JCAM2008} \cite{GaoM-2008}.

Besides travelling wave solutions and variable separated solutions,
the $3+1$ dimensional Jimbo-Miwa equation \eqref{eq:JM:pma-taipei-JM-2008} has
the following class of polynomial solutions:
\begin{equation} u_1=
 a_0+a_1x+a_2y+a_3z+a_4t+a_5xz+a_6xt+a_7yz+\frac 32 a_5yt +a_8 zt,
\label{eq:polynomialsolution:pma-taipei-JM-2008}
\end{equation}
where $a_i,\ 0\le i\le 8,$ are arbitrary constants. These are all polynomial solutions found with Maple among the class of polynomial functions with individual degrees of the independent variables less than two.
 Letting $u=u_1$, the condition \eqref{eq:specialcondofBT:pma-taipei-JM-2008} becomes
\begin{equation} a_2v_{xx}+a_1v_{xy}=0,
\end{equation}
and thus, one can easily generate other solutions, for example,
\begin{equation}
u_2=u_1|_{a_2=0}+f(x,t)+h(z,t),\ u_3=u_1+g(y,z)+h(z,t), \end{equation}
where $u_1$ is defined by \eqref{eq:polynomialsolution:pma-taipei-JM-2008} and $f,g,h$ can be arbitrary functions.
Taking $f,g,h$ as polynomials in the indicated variables engender two other classes of polynomial solutions.

All the presented solutions, including
travelling wave solutions, variable separated solutions and polynomial solutions, show the remarkable richness of the solution space of the $3+1$ dimensional Jimbo-Miwa equation \eqref{eq:JM:pma-taipei-JM-2008}, though the equation itself is expected to be non-integrable since it has solutions which are not single-valued functions in the neighborhood of their singularity surfaces \cite{DorizziGRW-JMP1986}.
A similar diversity situation of exact solutions can be found in solution spaces of typical nonlinear wave equations, for instance,
the $1+1$ dimensional equations: the Korteweg-de Vries equation \cite{MaY-TAMS2005,AktosunM-IM2006},
 the Boussinesq equation \cite{LiMLZ-IP2007,MaLH-NA2008}, the nonlinear Schr\"odinger type equation \cite{LeeLP-CSF2004} and the Hirota-Satsuma coupled KdV equation \cite{TamMHW-JPSJ2000}, and the $2+1$ dimensional equations:
  the Kadomtsev-Petviashvili equation \cite{BiondiniK-JPA2003},
the Davey-Stewartson equation \cite{HietarintaH-PLA1990,LouL-JPA1996} and
the Boiti-Leon-Pempinelli dispersive long-wave system \cite{Ma-PLA2003}.

\footnotesize

\section*{Acknowledgements}
The work was supported in part by the Established Researcher Grant
and the CAS faculty development grant of the University of South
Florida, Chunhui Plan of the Ministry of Education of China, Wang
Kuancheng Foundation, Zhejiang Normal University and Academia Sinica
in Taipei.

\end{document}